\documentclass[floatfix,twocolumn,showpacs,preprintnumbers,amsmath,amssymb,prl,superscriptaddress,longbibliography]{revtex4-1}

\usepackage{color}    
\usepackage{graphicx}
\usepackage{dcolumn}
\usepackage{bm}
\usepackage{subfigure}
\usepackage{amssymb}
\usepackage{multirow}
\usepackage{natbib}
\usepackage{amsmath}
\graphicspath{{plots/}}
\usepackage{hyperref}
\hypersetup{
    colorlinks = true,
    citecolor = blue
}
\usepackage[utf8]{inputenc}
\usepackage[T1]{fontenc}
\usepackage[normalem]{ulem}
	
\newcommand{\br}{\bm{r}}

\newcommand{\angstrom}{\mbox{\normalfont\AA}}
\renewcommand{\vec}[1]{\mathbf{#1}}

\begin{document}

\title{Thermal Signals from Collective Electronic Excitations in Inhomogeneous Warm Dense Matter}

\author{Zh.~A. Moldabekov} 
\email{z.moldabekov@hzdr.de}
\affiliation{Center for Advanced Systems Understanding (CASUS), D-02826 G\"orlitz, Germany}
\affiliation{Helmholtz-Zentrum Dresden-Rossendorf (HZDR), D-01328 Dresden, Germany}

\author{T. Dornheim} 
\affiliation{Center for Advanced Systems Understanding (CASUS), D-02826 G\"orlitz, Germany}
\affiliation{Helmholtz-Zentrum Dresden-Rossendorf (HZDR), D-01328 Dresden, Germany}

\author{A. Cangi}
\email{a.cangi@hzdr.de}
\affiliation{Center for Advanced Systems Understanding (CASUS), D-02826 G\"orlitz, Germany}
\affiliation{Helmholtz-Zentrum Dresden-Rossendorf (HZDR), D-01328 Dresden, Germany}

\begin{abstract}
We predict the emergence of novel collective electronic excitations in warm dense matter with an inhomogeneous electronic structure based on first-principles calculations. The emerging modes are controlled by the imposed perturbation amplitude. They include satellite signals around the standard plasmon feature, transformation of plasmons to optical modes, and double-plasmon modes. Most importantly, these modes exhibit a pronounced dependence on the temperature. This makes them potentially invaluable for the diagnostics of plasma parameters in the warm dense matter regime.
We demonstrate that these modes can be probed with present experimental techniques. 
\end{abstract}

\maketitle

\textit{Introduction.~} Collective excitations are ubiquitous in physics and emerge due to many-body interactions, for example, in quantum mechanics~\cite{mahan1990many}. 
An emerging research area where collective electronic excitations -- such as plasmons~\cite{bohm-pines-3} -- play a central role is warm dense matter (WDM)~\cite{graziani-book, Fortov2016, doe-report-17}. Research on WDM has received increased attention due to its relevance for fusion energy experiments, novel materials discovery, and high-energy astrophysical phenomena. From a technological point of view, warm dense conditions occur in the heating process of inertial confinement fusion capsules~\cite{KDFMLWG11,BH16}. In terms of materials science, understanding WDM propels the discovery of unexplored material properties such as novel chemistry~\cite{TRB08,Brongersma2015}, non-equilibrium effects~\cite{PHK06,EHH09} and phase transitions~\cite{KSM07}.
On a fundamental science level, probing warm dense conditions is essential for obtaining new insights on astrophysical objects, such as Earth's core~\cite{AGP99,NH04}, the interior of both solar planets~\cite{Militzer_2008,KDBLCSMR15,manuel} and exoplanets~\cite{NFKR11,KKNF12}, the properties of brown and white dwarfs~\cite{CBFS00, saumon1}, and neutron stars~\cite{Daligault_2009}. 

One labels the state of matter as WDM when (1) the Fermi energy $E_F$ and characteristic thermal energy of electrons $k_BT$ have the same order of magnitude, i.e., $\theta=k_BT/E_F\sim1$ and (2) electrons are strongly correlated, i.e., $r_s=a/a_B\geq 1$ where $a$ denotes mean interelectronic distance and $a_B$ the first Bohr radius~\cite{moldabekov_pre_18}. Additionally, WDM phenomena are transient in experiments -- covering the range from femtoseconds to picoseconds. 
Studying WDM phenomena is therefore highly challenging~\cite{graziani-book,new_POP} -- common concepts from well-established fields of physics do not apply: the persistence of quantum effects and strong coupling renders plasma physics concepts inaccurate, while accurate approaches from condensed-matter physics are often infeasible due to the length, time, and temperature scales involved. 

The aforementioned applications and the intricate nature of WDM have triggered a growing number of cutting-edge experimental activities, often in large-scale research facilities~\cite{MBRKA09, Fletcher2015, LCLS_2016, GFGN16, tschentscher_photon_2017, POP_exp_scat}. 
The array of experimental techniques that probe collective excitations in WDM include X-Ray absorption spectroscopy~\cite{VCCE2012:creation}, emission spectroscopy~\cite{PhysRevE.72.036408,PhysRevLett.109.065002}, X-Ray Thomson scattering~\cite{GGLR2003:demonstration}, resonant inelastic X-Ray scattering~\cite{HMvG2020:probing}, and the most recently developed ultrafast multi-cycle terahertz measurement technique~\cite{CCZT2021:ultrafast}. 

Due to its challenging nature, a reliable diagnostics of WDM properties (such as temperature, density) can only be achieved by joining efforts from experimental campaigns with accurate, first-principles modeling techniques~\cite{siegfried_review,kraus_xrts}. For example, consider determining the electronic temperature in warm dense samples. In principle, the temperature can be obtained from the detailed balance of the dynamic structure factor~\cite{So2010} that is probed in X-Ray Thomson scattering. In practice, however, the noise in the measured signals render a purely experimental temperature determination inaccurate~\cite{DLLN2009:temperature}.

Historically, WDM modeling has relied on dielectric models such as the random-phase approximation~\cite{ISI:A1954XZ29300001}. A formally exact formulation of dielectric models is provided by the local field correction~\cite{Ichimaru_RevModPhys}.
Their accurate parametrization is provided by quantum Monte-Carlo calculations of the uniform electron gas in its ground state~\cite{CA1980:ground,PhysRevLett.75.689} and at finite temperature ~\cite{BCDC13, DGSMFB16}. A notable example is the recently introduced effective static approximation~\cite{PhysRevLett.125.235001,Dornheim_PRB_2021}.
Extensions of dielectric models to take into account electron-ion collisions are based on the Mermin approach (MA)~\cite{mermin1970lindhard,Reinholz_2000,Fortmann_2010}, but only in an approximate manner.
An alternative simulation method for WDM diagnostics is Kohn-Sham density functional theory (KS-DFT)~\cite{hohenberg-kohn,KS65} and its extension to the time domain~\cite{runge-gross}. Similar to dielectric models, the electron-electron correlation is approximated in practice, but electron-ion collisions are tackled more directly. These calculations have become the workhorse among modern simulation methods in materials~\cite{PGB2015:dft, LBBB2016:reproducibility} and most recently also for WDM modeling~\cite{KRDM2008:complex, BSDH2016:xray, RCDB2021:firstprinciplesa}.

In prior WDM diagnostics efforts, homogeneous samples are commonly assumed, particularly regarding the numerical modeling aspects. Under this assumption, the electronic temperature of the induced WDM states has been determined from X-Ray Thomson scattering signals, for instance, in warm dense Aluminum either by dielectric models~\cite{Sperling_PRL_2015} or KS-DFT calculations~\cite{witte_prl_17,RCDB2021:firstprinciplesa}. Despite the combined efforts of experiment and simulation, determining the electronic temperature is still subject to significant uncertainties~\cite{MFKZ2018:firstprinciples}. 
Assuming a homogeneous WDM sample is justified when relaxation effects are considered. Their time scale is typically on the order of several picoseconds~\cite{PhysRevLett.110.065001, moldabekov_pre_18}. 
Moreover, this assumption underpins the uniform electron gas~\cite{CA1980:ground,DGSMFB16} as a paradigm for modeling WDM phenomena.

In this Letter, we break away from the common assumption of a homogeneous electronic structure in WDM diagnostics. We (1) predict the emergence of yet unexplored modes of collective excitations in \emph{inhomogeneous} WDM and (2) demonstrate the utility of these modes for the temperature diagnostics of WDM.

We show that features in the new collective modes are absent in the spectrum of homogeneous WDM states which is dominated by the well-known plasmon mode. The variety of features we observe depends on the degree of the density deviation from the homogeneous state. They include satellite signals around the standard plasmon feature, transformation of plasmons to optical modes, and double-plasmon modes.
We demonstrate that both the realization of the required perturbation amplitudes and the observation of the proposed features are feasible using present and upcoming experimental facilities~\cite{Sawada, RSI2020, PhysRevX.8.031068, Ofori_Okai_2018, Fletcher2015}. For example, the dynamics of emerging electronic structures was recently shown in laser-driven WDM samples where inhomogeneities are imposed by a periodic grating structure~\cite{PhysRevX.8.031068}. 
Furthermore, spatially modulating electronic structures can also be triggered by free electron lasers~\cite{Fletcher2015}, VUV lasers~\cite{Zastrau}, and THz lasers~\cite{Ofori_Okai_2018}.
Most importantly, we demonstrate that our proposed features exhibit a distinct temperature dependence and are thus uniquely suited for setting new standards of WDM diagnostics. 

\textit{Results.~} 
Our investigation begins with imposing spatial modulations on the warm dense uniform electron gas (UEG) by the Hamiltonian
\begin{align}\label{eq:ham}
H=H_{\rm UEG}+ \sum_{i=1}^N 2 U_0 \cos\left(\br_i \cdot \vec Q \right)\,,
\end{align}
where $N$ is the total number of electrons and $H_{\rm UEG}$ is the standard Hamiltonian of the UEG. Note that here and below we adopt Hartree atomic units, where the length is expressed in Bohr and the energy in Hartree.
This elementary Hamiltonian captures the fundamental physics that leads to the proposed collective mode. Likewise, it is motivated by the fact that electronic oscillations in WDM are well described by the UEG due to the relatively weak electron-ion coupling~ \cite{PhysRevLett.125.235001, CPP2015, GRABOWSKI2020100905}.
In Eq.~(\ref{eq:ham}), the degree of inhomogeneity is determined by the perturbation amplitude $U_0$. The length scale of the imposed modulations is set by the wave vector $\vec Q$. This Hamiltonian has been used extensively in prior work to study response properties of ambient and warm dense electrons with respect to local field corrections~\cite{PhysRevLett.69.1837,PhysRevLett.75.689} and response functions~\cite{dornheim_physrep_18,PhysRevLett.125.085001}.
\begin{figure}[t!]
\vspace{5mm}
\centering
\includegraphics[width=0.45 \textwidth]{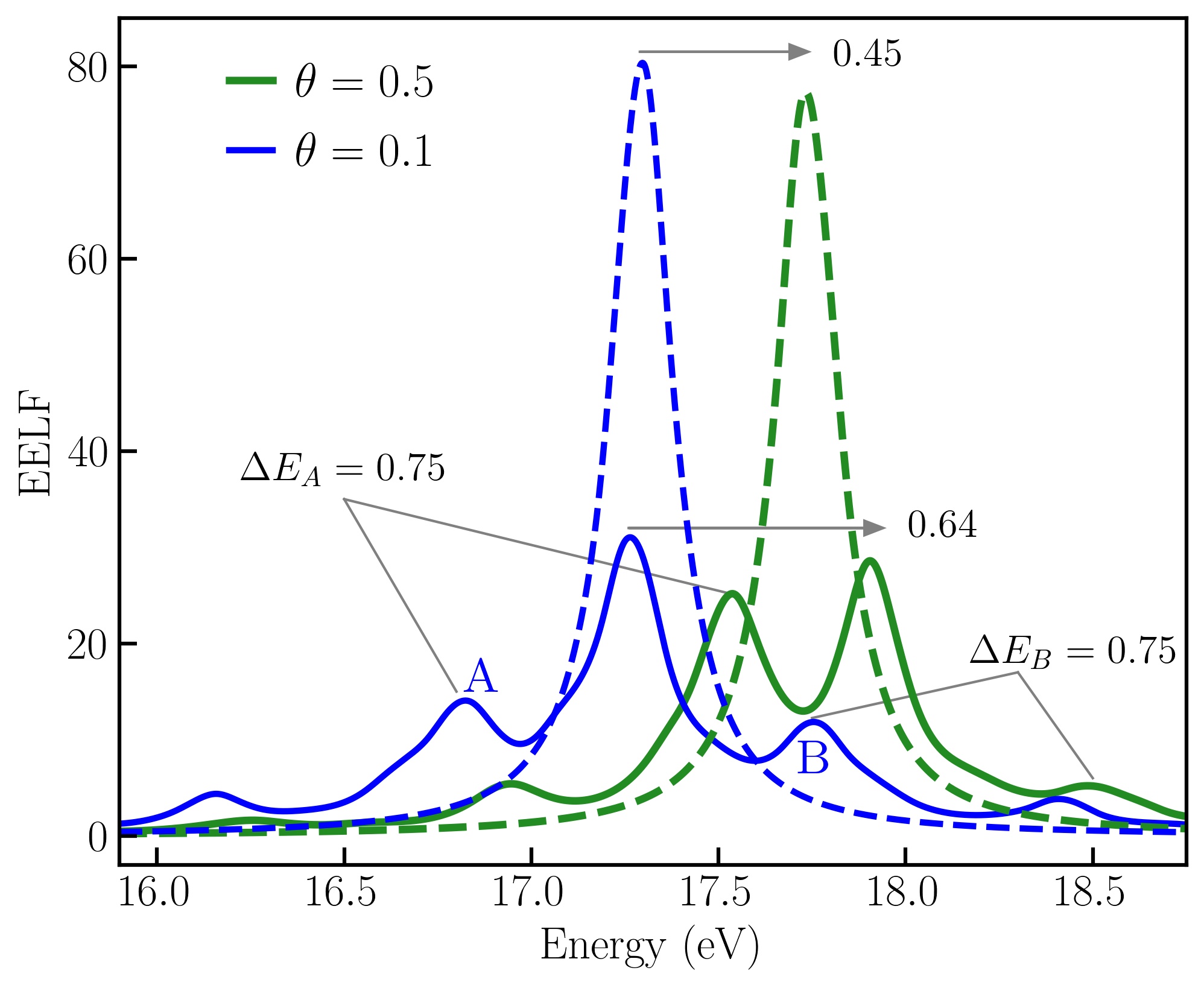}
\caption{\label{fig:1}
A novel mode of collective electronic excitation (solid blue and green curves) that emerges in inhomogeneous WDM. The new mode is observed in the EELF along the wave vector $\vec Q$ when a perturbation amplitude $U_0=0.1$ is imposed on the electronic structure. The EELF is shown for $\vec q=0.431~{\rm \angstrom}^{-1}$ at temperatures $\theta=0.5$ (solid blue) and $\theta=0.1$ (solid green). The new mode differs significantly from the well-explored plasmon mode (dashed blue and green curves). It also has a pronounced temperature dependence. 
}
\end{figure} 

We utilize the electron energy-loss function (EELF) to illustrate the emergence of collective electronic excitations in inhomogeneous WDM 
\begin{equation}
    \mathcal{L}(\vec q,\omega)=-{\rm Im}\left[\frac{1}{\varepsilon(\vec q, \omega)}\right]\,,
\end{equation}
which is defined in terms of the macroscopic dielectric function $\varepsilon(\vec q, \omega)$~\cite{PhysRevB.83.245122}.
Collective electronic excitations are identified by distinct features located at the maxima of the EELF~\cite{Ashcroft}. 
The EELF is proportional to the double-differential cross section in inelastic electron scattering. It is measured directly in electron energy-loss spectroscopy~\cite{Egerton_2008}. Likewise, the X-Ray Thomson scattering signals are proportional to the EELF via the dynamical structure factor~\cite{siegfried_review} obtained from the fluctuation-dissipation theorem~\cite{quantum_theory}. The collective electronic excitations we predict are computed using first-principles calculations of the EELF based on the Hamiltonian in Eq.~(\ref{eq:ham}). We use time-dependent DFT~\cite{runge-gross} as implemented in the GPAW code~\cite{GPAW1, GPAW2, ase-paper, ase-paper2} and employ the adiabatic local density approximation~\cite{Perdew_LDA}. We corroborate the accuracy of our results with quantum Monte-Carlo calculations. Details are provided in the Supplemental Material~\cite{supplement}. 

In the following, we consider the typical range of WDM conditions~\cite{Fortov2016} with a density parameter $r_s=2$ (corresponding to an electron number density $n\simeq 2\times 10^{23}~{\rm cm}^{-3}$) and a degeneracy parameter $\theta$ in the range from $0.1$ to $1.0$. The choice of amplitude and wave number of the perturbation is grounded in our interest of inhomogeneities on collective excitations which become relevant when $q/q_F\ll 1$~\cite{Hamann_cpp}.  
We therefore consider $U_0$ in the range from $0.1$ to $1.0$ and $ Q\simeq 0.84 q_F=1.5~ {\rm \angstrom}^{-1}$. 
Furthermore, these perturbation parameters are viable in current experimental setups~\cite{PhysRevX.8.031068, PhysRevLett.125.085001}. For example, available ${\rm THz}$ lasers with an intensity of $600~{\rm kV/cm}$~\cite{Ofori_Okai_2018} correspond to a perturbation amplitude $U_0\simeq 0.3$~\cite{PhysRevLett.125.085001}. Likewise, free electron lasers with intensities of up to $10^{22}~{\rm W/cm^2}$~\cite{Fletcher2015} provide perturbation amplitudes up to $U_0\approx 2$.
Alternatively, periodically inhomogeneous electronic structures can be generated using laser irradiation of pre-designed grating targets~\cite{PhysRevX.8.031068}.

Our central result is illustrated in Fig.~\ref{fig:1}. It displays (1) a new mode of collective electronic excitation (solid blue and solid green) that emerges in inhomogeneous WDM due to the perturbation in Eq.~(\ref{eq:ham}). Additionally, the figure displays (2) the pronounced temperature dependence of the new collective mode (solid blue versus solid green). In Fig.~\ref{fig:1}, the EELF is shown for $\vec q=0.431~{\rm \angstrom}^{-1}$. The new predicted mode emerges in the direction along the wave vector $\vec Q$. It is shown at two temperatures corresponding to $\theta=0.5$ (solid blue) and $\theta=0.1$ (solid green). It exhibits a rich structure distinctly different from well-known plasmon mode (dashed blue and dashed green). The new mode exhibits a dampened structure close to the position of the plasmon peak and shows several satellite peaks. More importantly, the new mode has a strong temperature dependence. The broadended feature in the new mode is shifted by about $0.64~{\rm eV}$ when the temperature is increased, while the plasmon mode is shifted by about $0.45~{\rm eV}$. The shift in the satellite peaks (${\rm A}$ and ${\rm B}$) is stronger, roughly $\Delta E_A \approx 0.75~{\rm eV}$ and $\Delta E_B \approx 0.75~{\rm eV}$.  Besides their energy shift, we expect the emergence of satellite peaks to be potentially helpful for diagnostics.

\begin{figure}[t!]
\centering
\includegraphics[width=0.49 \textwidth]{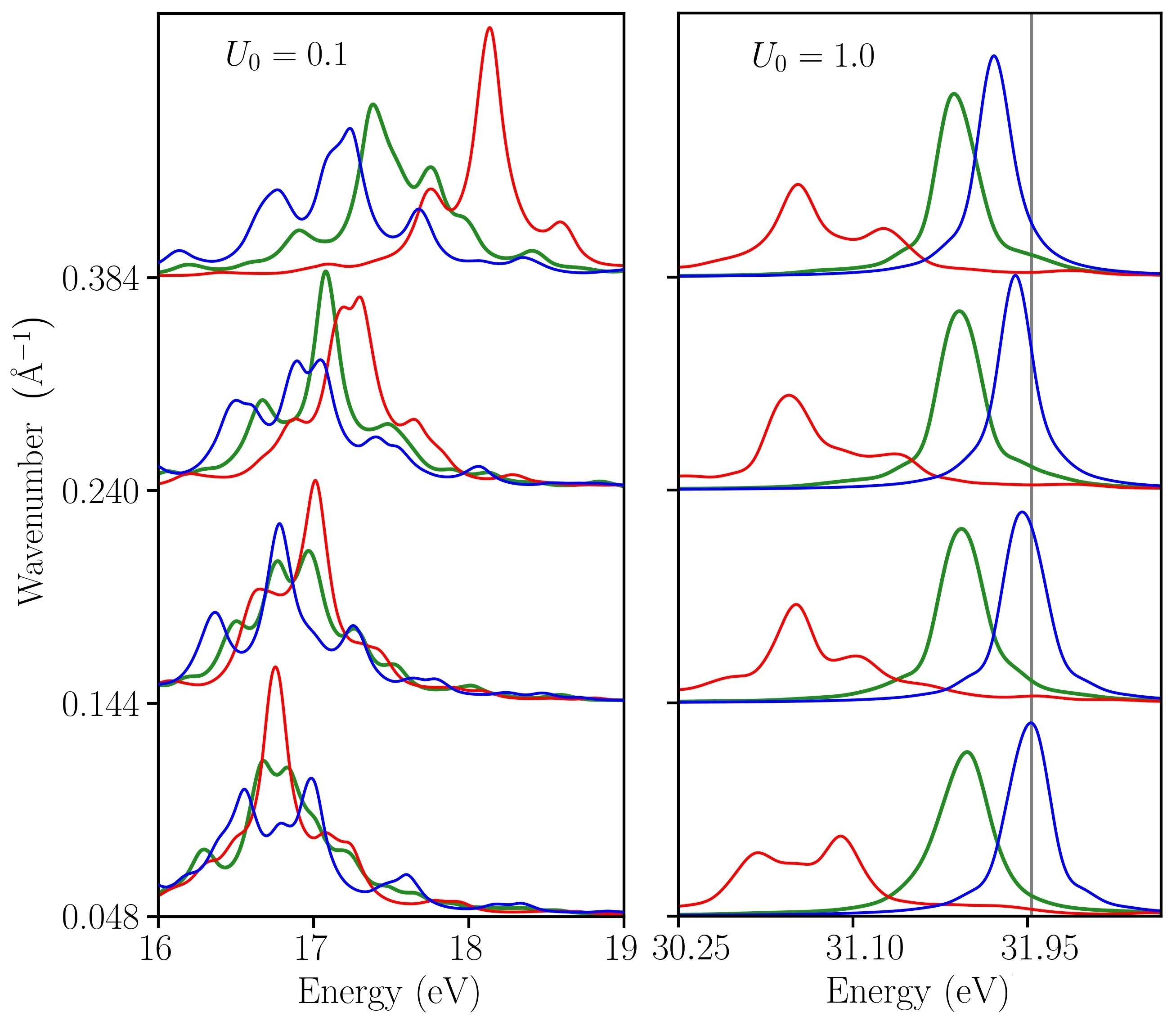}
\caption{\label{fig:2}
Prediction of novel collective modes in the EELF at weak and strong perturbation amplitudes. An optical mode emerges at a strong perturbation amplitude $U_0=1.0$. The dependence of these modes on $q$ and on the degeneracy parameter $\theta$ (solid blue denotes $\theta=0.1$, solid green $\theta=0.5$, and solid red $\theta=1.0$) is strong and depends on the imposed perturbation amplitude $U_0$. The vertical grey line has no significance. It is a guide to the eye and helps to observe that at $U_0=1.0$ the dispersion shifts to lower frequencies with increasing wave vector $q$.}
\end{figure} 
%

Next in Fig.~\ref{fig:2}, we predict additional collective modes and features that are observed along the wave vector $\vec Q$ for various values of $\vec q$. These features emerge when we assess a wider range of perturbation amplitudes. We consider both a weak perturbation amplitude $U_0=0.1$ and a strong amplitude $U_0=1.0$. We simultaneously probe the temperature dependence of these new modes for increasing values of the degeneracy parameter, $\theta=0.1$ (temperature $T\simeq 1.25~{\rm eV}$, solid blue), $\theta=0.5$ ($T\simeq 6.27~{\rm eV}$, solid green), and $\theta=1.0$ ($T\simeq 12.54~{\rm eV}$, solid red).

First (left panel of Fig.~\ref{fig:2}), we assess the emerging mode when the perturbation is weak ($U_0=0.1$). At low temperature ($\theta=0.1$), the mode remains centered around the same frequency, but its satellite features become more separated with increasing $\vec q$. With increasing temperature the mode peak intensity also rises and shifts to higher energies. Conversely, the satellite peaks become increasingly suppressed. This is due to thermal excitations that decrease the effect of the  inhomogeneity. 

Next in the right panel of Fig.~(\ref{fig:2}), we further increase the perturbation amplitude (to $U_0=1.0$). This leads to a transformation of the plasmon mode to an \textit{optical mode}.
The physics behind this transformation is a localization effect. Due to the strong perturbation, a large number of electrons becomes localized in the strong density enhancement regions. This is a manifestation of electron density oscillations around the center of mass of these localized regions. We observe that the energy of the \textit{optical mode} strongly depends on temperature. For example, at the smallest considered wave number $q = 0.048~\angstrom$, the energy of the \textit{optical mode} shifts by $0.4~{\rm eV}$ from $32~{\rm eV}$ at $\theta=0.1$ to $31.6~{\rm eV}$ at $\theta=0.5$. A further increase of the degeneracy parameter to $\theta=1.0$ results in a significant shift of the \textit{optical mode} to $\lesssim 31~{\rm eV}$. 

\begin{figure}[t]
\centering
\includegraphics[width=0.49 \textwidth]{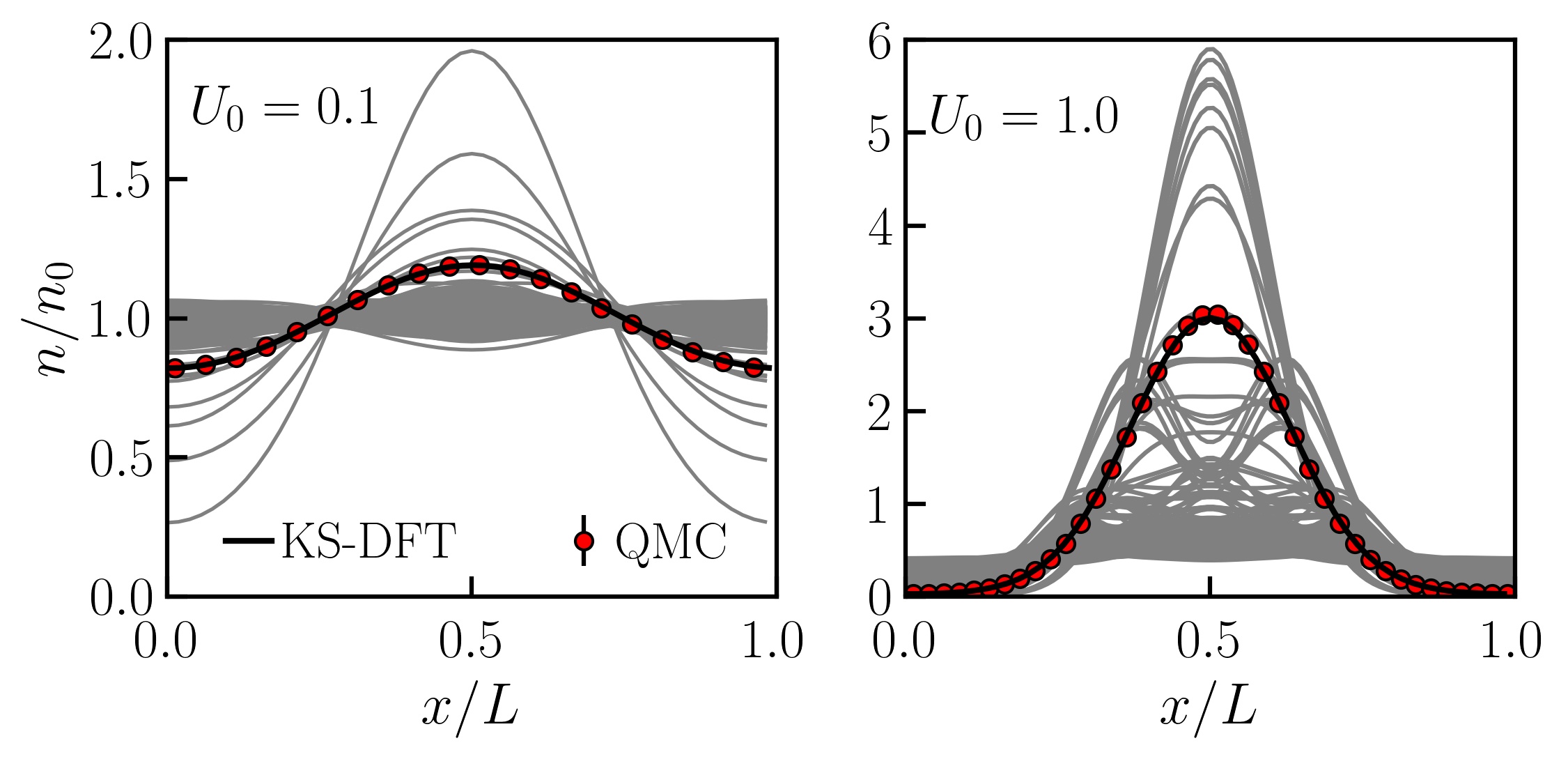}
\caption{\label{fig:3}
Analysis of the electronic structure for a weak ($U_0=0.1$) and strong ($U_0=1.0$) perturbation amplitude at a degeneracy parameter $\theta=1.0$. The orbital density (solid grey) and the total density (solid black) distributions from KS-DFT calculations provide an explanation for the emergence of the collective modes and features illustrated in Fig.~\ref{fig:2}. For illustrative purposes, the orbital density is scaled by the number of electrons. Furthermore, the comparison with quantum Monte-Carlo calculations (red circles) demonstrates the accuracy of our KS-DFT results. }
\end{figure}

Next, we provide further supporting evidence on the observation of the modes and features predicted in Fig.~\ref{fig:2}. To that end, we provide a detailed analysis of the underlying electronic structure in Fig.~\ref{fig:3}. There, we display the density distribution of orbital densities $n_i(\vec r)=|\phi_i(\vec r)|$ (solid grey) and the total density distribution $n(\vec r)=\sum_i f_i |\phi_i(\vec r)|^2$ (solid black) along $\vec Q$ at $\theta=1.0$, where $f_i$ denotes the occupation number. The left panel shows the results for a weak ($U_0=0.1$), the right panel for a strong ($U_0=1.0$) perturbation amplitude. While the KS orbitals $\phi_i(\vec r)$ are, strictly speaking, auxiliary quantities, they can nevertheless be used to explain the observed behavior of the EELF qualitatively. Furthermore, we confirm the accuracy of our KS-DFT calculations by comparing them to accurate quantum Monte-Carlo calculations based on the Hamiltonian in Eq.~(\ref{eq:ham})~\cite{supplement}. As shown, the total density distribution from KS-DFT (solid black) agrees very well with the quantum Monte-Carlo results (red circles) throughout. 
At $U_0=0.1$, most of the orbitals are spread across the simulation domain and several orbitals show a significant deviation from uniformity. This leads to the formation of satellite peaks around the plasmon peak observed in the left panel of Fig.~\ref{fig:2}. 
At $U_0=1.0$, most of the orbitals are localized at the central region.
This causes the formation of an optical mode due to the collective oscillation of these orbitals around the center of the localization region. Further increase in $U_0$ leads to a stronger signal from optical modes.     
%

So far, we discussed emerging modes along the direction of $\vec Q$. However, in the transpose direction to $\vec Q$, the EELF of the inhomogeneous system exhibits a mode similar to the standard plasmon~\cite{supplement}. 

\begin{figure}[t]
\centering
\includegraphics[width=0.45\textwidth]{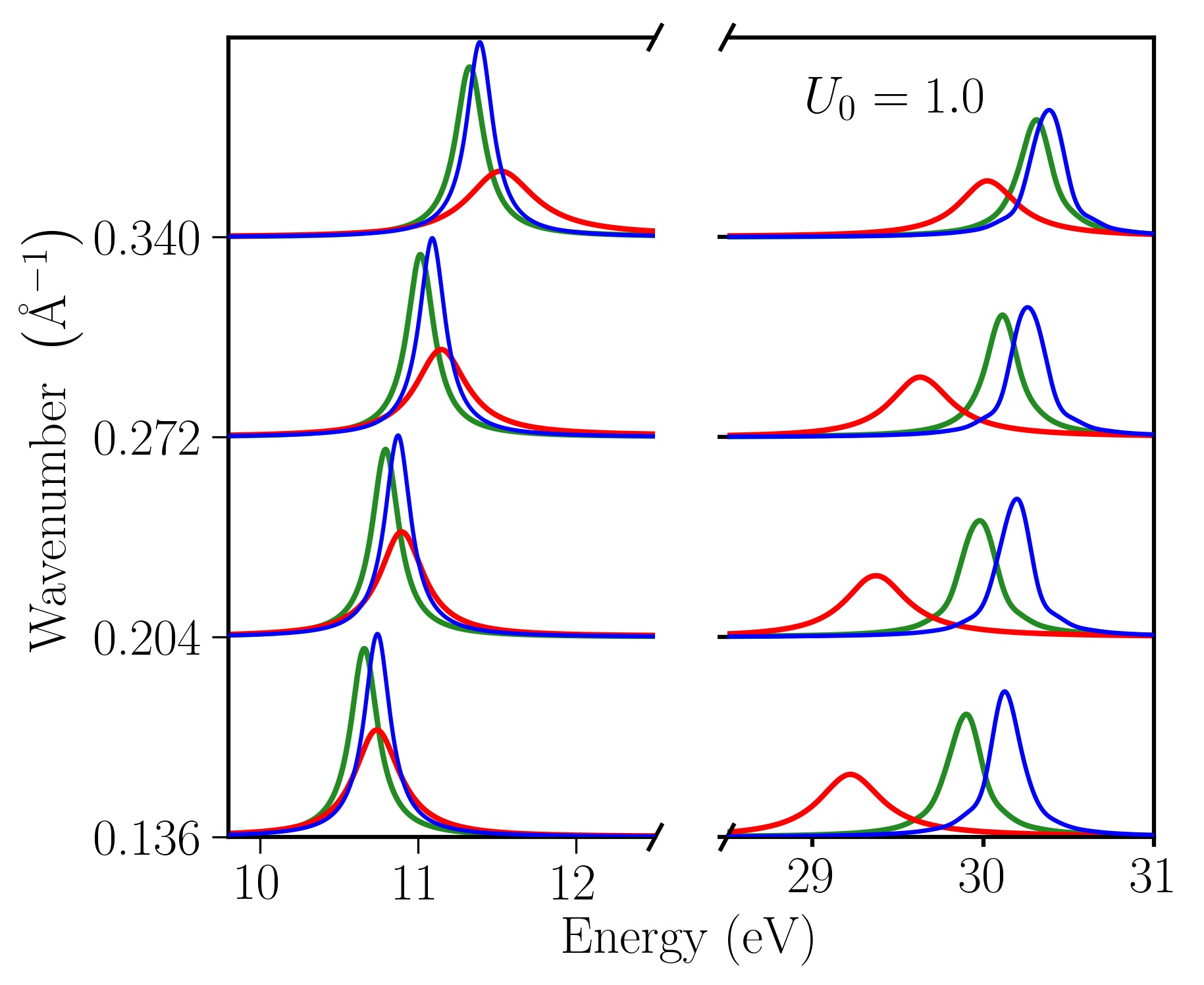}
\caption{\label{fig:4}
The emergence of a double-plasmon mode is observed at an angle of 45 degrees to $\vec Q$. The EELF is shown at a perturbation amplitude $U_0=1.0$ and for increasing temperature where solid blue denotes $\theta=0.1$, solid green $\theta=0.5$, and solid red $\theta=1.0$. The double-plasmon feature exhibits a pronounced dependence on the temperature.
}
\end{figure}

Finally in Fig.~\ref{fig:4}, we demonstrate the formation of a double-plasmon mode at an angle of 45 degrees to $\vec Q$.
The EELF is plotted for a perturbation amplitude of $U_0=1.0$ at various temperatures. The double plasmon signal appears at well separated energies. For example, at $q=0.136~\angstrom$ one signal emerges around $10.7~{\rm eV}$ and the other in the range from $29.1~{\rm eV}$ to $30.1~{\rm eV}$. 
The plasmon excitation at larger energies is particularly sensitive to thermal effects. It shifts by about $0.9~{\rm eV}$ when $\theta$ increases from $0.1$ to $1.0$.    

\textit{Conclusion.~} 
In this Letter, we have predicted the emergence of novel collective excitations in WDM when an inhomogeneous electronic structure is imposed. From a physical perspective, the well-known plasmon is transformed into a more complicated mode with multiple peaks. Its details can be readily controlled via the imposed perturbation amplitude. The observed features become richer when the degree of inhomogeneity is increased. Ultimately, this increase results in both an optical mode and a double-plasmon feature. These novel physical effects are interesting in their own right, and the required perturbation amplitudes can be realized in experimental facilities~\cite{supplement}. We also show that the energy resolution of present experiments is sufficient to resolve the emerging modes and features~\cite{supplement}.

From a practical point of view, the presented EELF spectra exhibit a substantially stronger dependence on the electronic temperature than the standard plasmon of homogeneous WDM. Furthermore, their directional dependence and richer structure provide additional constraints on theory and simulation. This can be used to solve the long-standing problem of temperature diagnostics in WDM, which is rather unconstrained at present. 

We expect that the interplay of experiment and simulation in probing inhomogeneous WDM will potentially become an invaluable new method of diagnostics enabling the reliable inference of important plasma parameters.
Prospective work to confirm our predictions in WDM experiments is planned. 

\begin{acknowledgments}
\section*{Acknowledgments}
The authors acknowledge particularly useful discussions with Michael Bussmann.
This work was funded by the Center for Advanced Systems Understanding (CASUS) which is financed by Germany's Federal Ministry of Education and Research (BMBF) and by the Saxon Ministry for Science, Culture and Tourism (SMWK) with tax funds on the basis of the budget approved by the Saxon State Parliament.
We gratefully acknowledge CPU-time at the Norddeutscher Verbund f\"ur Hoch- und H\"ochstleistungsrechnen (HLRN) under grant shp00026 and on a Bull Cluster at the Center for Information Services and High Performace Computing (ZIH) at Technische Universit\"at Dresden.
\end{acknowledgments}

\bibliography{ref,mb-ref,ref_attila}

\end{document}